\documentclass[twocolumn]{aastex63}
\usepackage{amsmath}
\usepackage{color}

\def\beq{\begin{eqnarray}}
\def\eeq{\end{eqnarray}}

\received{}
\revised{}
\accepted{}

\submitjournal{ApJ}

\shorttitle{Nucleosynthesis in NDAFs and effects on galaxies}
\shortauthors{Liu et al.}

\begin{document}

\title{Neutrino-dominated accretion flows: second nucleosynthesis factory in core-collapse supernovae and regulation of iron markets in galaxies}

\correspondingauthor{Tong Liu}
\email{tongliu@xmu.edu.cn}

\author[0000-0001-8678-6291]{Tong Liu}
\affiliation{Department of Astronomy, Xiamen University, Xiamen, Fujian 361005, China}

\author{Yan-Qing Qi}
\affiliation{Department of Astronomy, Xiamen University, Xiamen, Fujian 361005, China}

\author[0000-0002-4223-2198]{Zhen-Yi Cai}
\affiliation{CAS Key Laboratory for Research in Galaxies and Cosmology, Department of Astronomy, University of Science and Technology of China, Hefei, Anhui 230026, China}
\affiliation{School of Astronomy and Space Science, University of Science and Technology of China, Hefei, Anhui 230026, China}

\author[0000-0002-0771-2153]{Mouyuan Sun}
\affiliation{Department of Astronomy, Xiamen University, Xiamen, Fujian 361005, China}

\author{Hui-Min Qu}
\affiliation{Department of Astronomy, Xiamen University, Xiamen, Fujian 361005, China}

\author{Cui-Ying Song}
\affiliation{Department of Astronomy, Xiamen University, Xiamen, Fujian 361005, China}
\affiliation{Tsung-Dao Lee Institute, Shanghai Jiao Tong University, Shanghai 200241, China}

\begin{abstract}
Cosmic metals are widely believed to be produced by supernovae (SNe) and compact object mergers. Here, we discuss the nucleosynthesis of neutrino-dominated accretion flows (NDAFs) with outflows in the core-collapse SNe (CCSNe), and show that the outflows from NDAFs can have a significant contribution to the $^{56}$Ni abundance in the faint explosions if the masses of the progenitor stars are within about $25-50$ $M_\odot$. Less massive progenitor stars can produce more $^{56}$Ni than their more massive counterparts in the NDAF outflow nucleosynthesis channel. Therefore, we find that the total (i.e., CCSNe and NDAF outflows) $^{56}$Ni mass per CCSN depends only weakly upon the mass of progenitor stars. In the metallicity evolution, the ratio of $^{56}$Fe (decayed by $^{56}$Ni) mass to the initial total gas mass can increase by $\sim$ 1.95 times if the upper limits of the nucleosynthesis yields from NDAF outflows and CCSNe are considered. Our results might have significant implication for chemical evolution of the the solar neighborhood, galaxies, and active galactic nuclei.
\end{abstract}


\section{Introduction}

Neutrino-dominated accretion flows (NDAFs) in the center of collapsars or compact object mergers are the plausible central engine of gamma-ray bursts \citep[GRBs, for reviews, see, e.g.,][]{Liu2017,Zhang2018}. Because NDAFs around black holes (BHs) with very high accretion rates ($\dot{M}\sim$ $0.001-10$ $M_\odot~\rm s^{-1}$) are in the state of high density ($\rho \sim 10^{10}-10^{13}~\rm g~cm^{-3}$) and temperature ($T \sim 10^{10}-10^{11}~\rm K$), photons are fully trapped and the neutrino-participation processes intensively occur in the disk; only neutrinos can escape from the disk surface to dissipate the viscous heating energy \citep[e.g.,][]{Popham1999,Narayan2001,Kohri2002,Kohri2005,Lee2005,Gu2006,Liu2007,Janiuk2007,Chen2007,Kawanaka2007,Xue2013}. Neutrino annihilations above or below the disk will drive ultra-relativistic jets to trigger GRBs \citep[e.g.,][]{Ruffert1997,Rosswog2003,Zalamea2011}.

In the collapsar scenario, the initial mass supply rates can keep the accretion processes in the NDAF phase; however, the jets are possibly choked in the envelopes of the collapsars, especially for the low-metallicity massive progenitor stars. Eventually, the jets might break out to power GRBs if the Blandford-Znajek (BZ) mechanism \citep{Blandford1977} dominates over the neutrino annihilations \citep[e.g.,][]{Nakauchi2013,Matsumoto2015,Liu2018,Nagataki2018}, unless the hydrogen envelope has been expelled prior to explosion. Still, NDAFs play primordial roles in at least five scenes.

First, most of neutrinos ($\sim 99 \%$) emitted from the disk do not participate in the annihilations but escape freely, so NDAFs are important sources of the MeV neutrinos after the explosion of core-collapse supernovae (CCSNe). Although the typical fluence is one to two orders of magnitude lower than that of CCSNe, the neutrinos of NDAFs in the Local Group ($\lesssim$ 1 Mpc) might be detected by the future liquid-scintillator detector Low Energy Neutrino Astronomy (LENA) and Hyper-Kamiokande \citep[e.g.,][]{Liu2016,Liu2017b,Wei2019}. Of course, the neutrinos from NDAFs in the center of CCSNe with different mass, metallicity, and initial explosion energy should contribute the neutrino background \citep{Wei2021}.

Second, the jet precession driven by an NDAF around a spinning BH \citep[e.g.,][]{Blackman1996,PortegiesZwart1999,vanPutten2003,Reynoso2006,Lei2007,Liu2010} or the anisotropic emission of neutrinos from NDAFs \citep[e.g.,][]{Suwa2009,Liu2017b} can release the gravitational waves (GWs) in $\sim 1-100~\rm Hz$ , which at the distance of 10 kpc, even 1 Mpc, might be detected by the Einstein Telescope (ET), the Decihertz Interferometer Gravitational Wave Observatory/Big Bang Observer (DECIGO/BBO), and ultimate-DECIGO \citep[e.g.,][]{Sun2012,Wei2020}. Detections of these neutrinos or GWs can confirm the existence of the NDAFs and constrain the mass and spin of the central BHs in the collapsar or merger scenario \citep[e.g.,][]{Liu2017b}.

Third, NDAFs feed the central BHs and significantly alter their masses and spins if the outflows are inefficient \citep[e.g.,][]{Janiuk2008,Song2015}. For example, by using the fall-free approximation (corresponding to the very faint explosion), the initial mass-supply rates can be estimated as $\sim 1~ M_\odot~\rm s^{-1}$ and the accretion process can last about 10 s for the progenitors of $\sim 40~M_\odot$; although the physical supply rate might be lower than $1~ M_\odot~\rm s^{-1}$, the central BHs, $\sim 5~M_\odot$, will grow by several solar masses. The stellar-mass BHs should undergo the fallback hyperaccretion history with different explosion energy when they are just born in the collapsars even be immediately kicked from the centers \citep[e.g.,][]{Zhang2008}. It provides a possible way to understand the ``first (or lower) mass gap'' ($\sim 2.5-5~M_\odot$) in the stellar-mass BH distribution \citep{Liu2021}. In addition, the BH spins might also be significantly changed via the hyperaccretion processes. Note that the relative importance of the inflows and outflows can affect the lightcurves and luminosities of GRBs and CCSNe \citep[or kilonovae, e.g.,][]{Liu2017,Song2018,Song2019}.

Fourth, strong outflows from the BH hyperaccretion systems should continuously inject and resupply gas into the envelope of collapsars, increase the accretion timescale and induce fluctuations in the accretion rates \citep{Liu2019}. This mechanism can explain the unusually bright, long-lived iPTF14hls \citep[e.g.,][]{Arcavi2017} and some supernovae (SNe) with double-peak lightcurves \citep[e.g.,][]{Mazzali2008}.

Fifth, a mass of free protons and neutrons abounds in the NDAFs (especially in the inner regions) and the cooling processes in the outflows should synthesize abundant heavy metals. The synthesis products from the NDAF outflows are quite different between the relatively proton-rich circumstances in the collapsars and the neutron-rich condition in the compact object mergers \citep[e.g.,][]{Surman2006,Liu2013,Xue2013,Janiuk2014,Siegel2017}. In this work, we do not consider the compact object mergers, and the neutron-rich condition is irrelevant to our study. Actually, for the BH-NDAF systems in the center of the collapsars, the difference of the metallicity of the progenitor stars, i.e., the electron fraction $Y_{\rm e}$ at the outer boundary, can affect the sorts and yields of the metals from the NDAF outflows \citep[e.g.,][]{Pruet2004,Surman2005,Surman2006,Liu2013,Liu2017b,Xue2013,Janiuk2014,Song2019}.

Type Ia SNe are the thermonuclear explosions originated from the accretion white dwarfs (WDs) in the close binaries or the double-WD mergers. Some of them are considered as ``standard candles'' to determine the cosmological parameters. They are also believed to be one of the most important nucleosynthesis factories to produce heavy metals including the iron group \citep[see, e.g.,][]{Woosley1986,Arnett1996, Hoflich1998,Hillebrandt2000}.

Massive stars ($\gtrsim 8~M_\odot$) undergoing core-collapse at the end of their lives can trigger CCSNe (and Hypernovae). The nucleosynthesis (especially $^{56}$Ni) processes in such energetic SNe have been widely studied \citep[e.g.,][and references therein]{Woosley1986b,Woosley1995,Nakamura2001,Woosley2002,Heger2003,Maeda2003,Fryer2004,Nomoto2006,Fujimoto2007,Maeder2009,Heger2010,Winteler2012,Sukhbold2016,Mosta2018,Kobayashi2020}. It is worth mentioning that the rich $^{56}$Ni might be produced in the collapsars powering GRBs \citep[e.g.,][]{Woosley1999}, e.g., GRB 980425 associated with SN 1998BW \citep[e.g.,][]{Woosley1999a,Sollerman2000}. In order to explain the observations of SN 1998bw and SN 2003dh, \citet{Pruet2004} firstly investigated the nucleosynthesis in outflows from the inner region of the accretion disk formed after the collapse of a rotating massive star. For more massive stars ($\gtrsim 25~M_\odot$), BHs should generally be born in their center, which will lead the hyperaccretion processes. All types of SNe are profoundly crucial to the chemical evolution in galaxies and active galactic nuclei \citep[AGNs; see, e.g.,][]{Barth2003,Dietrich2003,Maiolino2003,Kobayashi2006,Kobayashi2020,Maiolino2019}. There are still some crisis on the metal abundance, such as in the solar neighborhood \citep[e.g.,][]{Kobayashi2020} and on the high-metallicity quasars in high redshift \citep[e.g.,][]{Onoue2020}.

In this paper, we focus on the $^{56}$Ni synthesis of the NDAFs with outflows in the CCSN scenarios for different progenitors and discuss their contribution to the chemical evolution of galaxies and AGNs. In Section 2, we briefly study the NDAFs with outflows and explore their nucleosynthesis conditions, then result the relations between the total $^{56}$Ni yields and progenitor masses. We briefly estimate the contributions of NDAF outflows and CCSNe to the $^{56}$Fe yields in the chemical evolution of galaxies in Section 3. Conclusions and discussion are made in Section 4.

\section{NDAFs with outflows}
\subsection{Model}

Here we present a simplified NDAF model in the presence of disk outflows. The relation between the accretion rate at any radius $\dot M (r)$ and at the outer boundary $\dot M_{\rm outer}$ can be described as a power law \citep[e.g.,][]{Blandford1999,Yuan2012,Yuan2014,Sadowski2015,Sun2019}, which is expressed as
\beq
\dot{M} (r) =\dot{M}_{\rm outer} \left(\frac{r}{r_{\rm outer}} \right)^{p},
\eeq
where $r_{\rm outer}$ is the outer boundary of the disk and can be determined by integrating the BH mass in the density profiles of the collapsar model \citep[e.g.,][]{Liu2018}. For a BH of $\sim 5~ M_\odot$ in the collapsar of the low-metallicity progenitor star of $\sim 40~ M_\odot$, $r_{\rm outer}$ is about $50 ~r_{\rm g}$, where $r_{\rm g}=GM_{\rm BH}/c^{2}$ is the Schwarzschild radius and $M_{\rm BH}$ is the mass of the BH. The index parameter $p$ determines the strength of the outflows, which is an uncertain parameter for NDAFs. We take the inner boundary of the disk $r_{\rm inner}\simeq r_{\rm ms}=(3+Z_{2}-\sqrt{(3-Z_{1})(3+Z_{1}+2Z_{2})})r_{\rm g}$, where $r_{\rm ms}$ is the marginally stable orbit radius, $Z_{1}=1+(1-a_*^{2})^{1/3}[(1+a_*)^{1/3}+(1-a_*)^{1/3}]$,
$Z_{2}=\sqrt{3a_*^{2}+Z_{1}^{2}}$, and $a_*$ ($0 < a_* <1$) is the dimensionless spin parameter of the BH \citep[e.g.,][]{Bardeen1972,Kato2008}.

We can calculate the structure of a steady and axisymmetric NDAF by considering the dynamic equations outlined in \citet{Liu2010} and \citet{Sun2012}.

The total pressure $P$ related to the local net accretion rate is the sum of contributions from four terms, i,e., the gas pressure, the radiation pressure, the electron degeneracy pressure, and the neutrino pressure \citep[e.g.,][]{Kohri2005,Liu2007},
\beq
P=P_{\rm gas} + P_{\rm rad} + P_{\rm e} + P_\nu,
\eeq
and the energy balance equation related to the local net accretion rate is,
\beq
Q_{\rm vis}^{+}= Q_{\rm adv}^{-}+ Q_{\rm photondis}^{-} + Q_{ \rm \nu }^{-},
\eeq
where $Q_{\rm vis}^{+}$, $Q_{\rm adv}^{-}$, $Q_{\rm photondis}^{-}$, and $Q_{\rm \nu}^{-}$ denote the viscous heating rate, and the cooling rates due to the advection, photodisintegration, and neutrino losses, respectively \citep[e.g.,][]{Kohri2005,Liu2007,Xue2013}. Here we ignore $Q_{\rm photondis}^{-}$ because it is much less than the neutrino cooling rate in the inner region of the disk \citep[e.g.,][]{Janiuk2004,Liu2007}. The detailed neutrino physics in the above two equations can be found in \citet{Liu2017}. It should be mentioned that the above two equations are all related to the local net accretion rate, so we consider that the equation of state is still satisfied and no cooling rate of the outflows is included in the energy equation. Thus the total energy dissipated by the outflows depends on the difference in the viscous heating rates between at the inner boundary and at the outer boundary. This solution should be better than the method of the parameterized entropy \citep[e.g.,][]{Surman2006,Miller2020} on the descriptions of the disk outflows.

We define the neutrino-cooling factor $f_\nu = Q_{\rm \nu}^{-}/Q_{\rm vis}^{+}$, as well as the advection factor $f_{\rm adv} = Q_{\rm adv}^{-}/Q_{\rm vis}^{+}$ \citep[e.g.,][]{Chen2007,Liu2017}. $f_\nu \geq 0.5$ is considered as the ignition condition for NDAFs. The main ingredients of the nucleosynthesis in the disk outflows including the initial density, temperature, and materials liable to synthesis, depend critically upon the state of the disk.

\subsection{Nucleosynthesis conditions}

\begin{figure}
\centering
\includegraphics[width=1.0\linewidth]{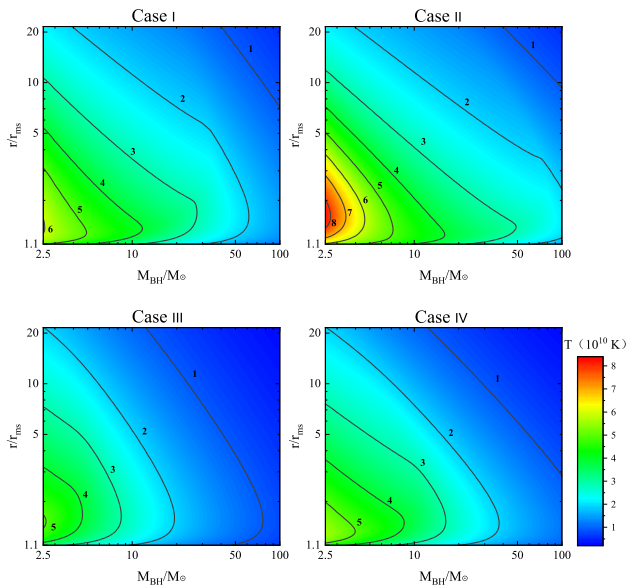}
\caption{Contours of the temperature $T/(10^{10}$ K) of the NDAFs with outflows on the $M_{\rm BH}-r$ planes for four cases. Cases I-IV correspond to ($\dot{m}_{\rm outer}$, $p$) = (1, 0.8), (1, 0.3), (0.1, 0.8), and (0.1, 0.3), respectively, where $\dot{m}_{\rm outer}$ = $\dot{M}_{\rm outer}$/($M_\odot~\rm s^{-1}$).}
\label{fig:contour_temperature}
\end{figure}

In the previous NDAF studies, the BH mass is often fixed to $3~M_\odot$. However, in the collapsar scenario, the mass of the newborn BH is related to the mass and metallicity of the progenitors \citep[e.g.,][]{Fryer1999,Heger2010}. Thus we firstly calculate the density, temperature, and neutrino-cooling factor of the NDAFs with different BH masses by fixing the viscous parameter of the disk $\alpha=0.1$ and the dimensionless spin parameter $a_* = 0.9$.

We then obtain the density profiles of the NDAFs with outflows for various $M_{\rm BH}$ and find that the disks are dense enough for the nucleosynthesis even at $r\simeq 50 r_{\rm g}$ (the requirement of the nucleosynthesis on the density is not very strict). The contours of the disk temperature $T/(10^{10}$ K) and the neutrino-cooling factor $f_\nu$ on the $M_{\rm BH}-r$ plane for four cases are shown in Figures 1 and 2, respectively; Cases I-IV correspond to ($\dot{m}_{\rm outer}$, $p$) = (1, 0.8), (1, 0.3), (0.1, 0.8), and (0.1, 0.3), respectively, where $\dot{m}_{\rm outer}$ = $\dot{M}_{\rm outer}$/($M_\odot~\rm s^{-1}$). Note that $p$ = 0.3 and 0.8 denote the weak and strong outflows from the disk, respectively.

As shown in Figure 1, the NDAF of Case II has the highest temperature since the corresponding accretion rate is the largest and the outflow is the weakest among the four cases. Nevertheless, in all cases, the temperature at $r \lesssim$ 20 $r_{\rm ms}$ is higher than $10^{10}$ K for $M_{\rm BH} \lesssim 100~M_\odot$. The initial temperature of the outflows is close to the disk temperature and is clearly high enough to trigger and maintain the nucleosynthesis processes. However, the NDAFs around the BHs with $M_{\mathrm{BH}}\gtrsim 50~M_\odot$ do not satisfy the ignition condition since their $f_\nu \leq 0.5$ (see Figure 2); that is, such NDAFs with outflows are not ideal for the nucleosynthesis.

In whichever cases of Figures 1 and 2, for the low mass BHs ($\sim 3-5 ~M_\odot$), as the products of the neutron star (NS)-NS or BH-NS mergers, the nucleosynthesis is efficient in the outflows from NDAFs, which can power the luminous kilonovae. We argue that their lightcurves are similar to the SNe with steep decay, named `quasi-SNe' \citep{Song2018}.

\subsection{Contribution on $^{56}$Ni yields}

\begin{figure}
\centering
\includegraphics[width=1.0\linewidth]{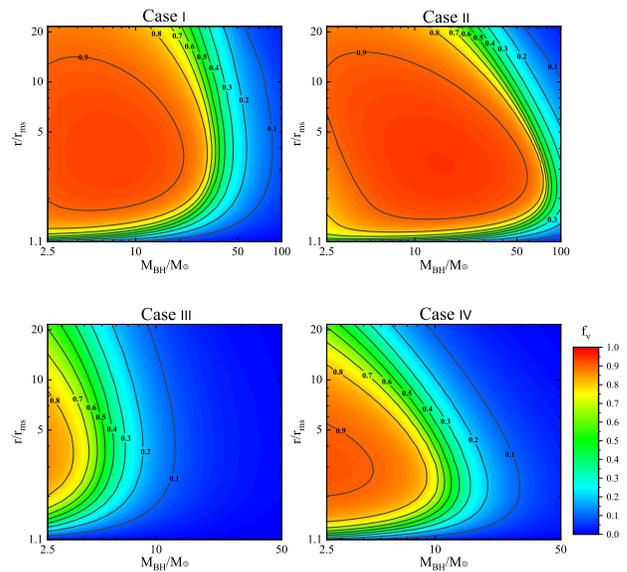}
\caption{Same as Figure~\ref{fig:contour_temperature}, but for contours of the neutrino-cooling factor $f_\nu$.}
\end{figure}

Since there is no unique link between the CCSN explosions and the central accretion processes in simulations, we adopt the initial explosions are very faint ($\sim 0$ B, where $1~\rm B = 10^{51}~ erg$) and very drastic ($\sim 100$ B) as two extreme cases, corresponding to the NDAFs with high accretion rates in the center of the collapsars and the violent CCSNe in the condition of [Mg/Fe] = 0 and [O/Fe] = 0, to describe the roughly upper limit of the total $^{56}$Ni yields.

In the collapsar scenario, based on the above solutions of different BH masses, we can estimate the approximately upper limits of the $^{56}$Ni yields of the outflows from the NDAFs. Since considering the effects of the BH masses on the density profiles of the progenitors in the fall-free approximation, the initial mass supply processes via fall back in the collapsars last about 10 s, and the rate is about 1 $M_\odot~\rm s^{-1}$ for the massive progenitors and decreases subsequently. In the following interval of about 50 s, the mean rate is about 0.1 $M_\odot~\rm s^{-1}$ \cite[e.g.,][]{Liu2018,Wei2019}. If the accretion rate at the outer boundary is assumed to equal the supply rate we can obtain the outflow mass in the condition of $f_\nu \geq 0.5$ with $p$ in the range of 0.3 to 0.8. The efficiency of the synthesis in the outflows is very high; in fact, almost all materials can be converted into the elements not lighter than $^4$He \citep[e.g.,][]{Surman2011,Xue2013}. \citet{Surman2011} considered that up to around 50$\%$ of the material ejected from BH accretion disks will become $^{56}$Ni under some optimal conditions. In contrast, \citet{Miller2020} found that very little $^{56}$Ni is ejected in their simulations. Here we assume that approximately 10$\%$ of the outflow materials are synthesized into $^{56}$Ni \citep[e.g.,][]{Surman2011,Song2019,Zenati2020}. We verify this assumption by using the code of the nuclear statistical equilibrium in proton-rich environments \citep[$Y_{\rm e} \sim 0.45-0.50$, see][]{Seitenzahl2008}. At the final step, we adopt the numerical results in \citet{Heger2002} to link the BH mass with the mass of the corresponding low-metallicity progenitor star. Here we follow \citet{Heger2002} to consider that the BHs are born in the center of the collapsars for the progenitor star masses $\gtrsim 25~M_\odot$.

\begin{figure}
\centering
\includegraphics[width=1.1\linewidth]{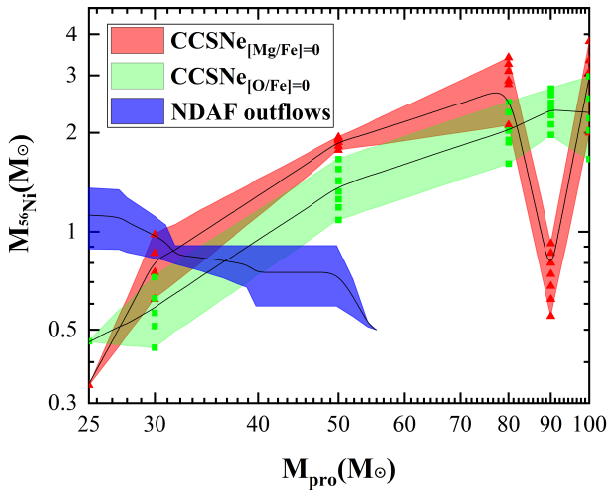}
\caption{Roughly upper limits of the $^{56}$Ni yields from the NDAFs with outflows and the CCSNe as functions of the progenitor star masses. The CCSN data are adopted from \citet{Umeda2008}. The black lines denote the medians of those regions.}
\end{figure}

Some stars might lose their partial envelopes owing to the binary interactions \citep[e.g.,][]{Podsiadlowski2003} or the strong winds \citep[e.g.,][]{Maeder1992,Heger2003}, then the very different outcomes are emerged at the end of their life. For example, the final core structure could be structurally changed even for the very massive progenitors in binary interactions, an NS rather than a BH might be born \citep[e.g.,][]{Podsiadlowski2003}. Moreover, rotations may be increasingly important to the massive stars \citep[e.g.,][]{Fryer2000,Fryer2004a,Heger2005,Maeder2012}, which also seriously affects the mass supply rate \citep[e.g.,][]{Liu2018}. Some massive stars should expel their hydrogen envelope prior to Ib/c SN explosion via winds or rotations \citep[e.g.,][]{Woosley1993,Yoon2005}. These above impacts and factors are not considered in this work. It should be mentioned that the NDAF contribution in nucleosynthesis is negligible at the stars within $\sim 8-25~M_\odot$. Because if the explosion energies of CCSNe for the stars, $\gtrsim 8~M_\odot$, are no much difference, the mass supply rates for the NSs owning the crusts should be much lower than these for the BHs \citep[e.g.,][]{Zhang2008,Liu2021}, then the fallback accretion is inefficiency for NSs. Additionally, for the newborn NSs in their centers, the strong magnetic fields will destroy the inner regions of the disks and prevent the accretion processes, then NDAFs can not be ignited. Thus the NSs will hold their long-lived fates and avoid collapse to the BHs unless the critically massive NSs spin down.

Figure 3 shows the roughly upper limits of the $^{56}$Ni yields from the NDAFs with outflows (blue shaded region) and the CCSNe (red and green shaded regions in the conditions of [Mg/Fe] = 0 and [O/Fe] = 0 and the explosion energy in the range of $1-100$ or $150 ~\rm B$) as functions of the progenitor star masses. The black lines denote the medians of those regions. The CCSN data are adopted from \citet{Umeda2008}, because the most metal-poor halo stars are believed that their abundances are satisfied with [Mg/Fe] $\geq$ 0 and [O/Fe] $\geq$ 0 according to observations. That means for [Mg/Fe] $<$ 0 or [O/Fe] $<$ 0, the $^{56}$Ni yields should be less than these shown in Figure 3 \citep{Umeda2008}. It should be mentioned that the shaded regions in Figure 3 just reflect the roughly upper limits of the $^{56}$Ni yields, or rather, the abilities of CCSNe and NDAFs on the $^{56}$Ni synthesis, because we set the terms for the fall-free approximation of the density profile, $p\geq 0.3$ of the NDAF outflows, and [Mg/Fe] = 0 and [O/Fe] = 0 of CCSNe. More importantly, contrary to the energetic CCSNe requiring the high explosion energy, the BH mass growth (i.e., fallback accretion) favors in the condition of the low explosion energy. In other words, there is the energy and matter competitions between the CCSN explosions and BH fallback accretions. Hence, Figure 3 indicates that the upper limits of the total $^{56}$Ni yields should depend weakly upon the explosion energy. The hypernova models can increase ten times or more $^{56}$Ni yields than the normal CCSNe \citep[e.g.,][]{Nomoto2004} and has been included to investigate the galactic evolution of the SN rates \citep[e.g.,][]{DeDonder2003}. In those energetic explosions, the fallback accretion is generally inefficient. As well as the hypernova models \citep[e.g.,][]{Kobayashi2020}, we consider that nucleosynthesis of the NDAF outflows in the center of the faint or failed CCSNe is another plausible way, which can explain the observations of the metal abundance in our neighborhood and the chemical evolution of galaxies. Moreover, \citet{Kobayashi2020} estimated that the proportion of the failed SNe is perhaps as large as 50$\%$ early in the history of the Galaxy, which might support the consequence of the NDAF outflows on synthesis.

It should be noted that the space below these regions is allowed for the NDAFs and CCSNe in Figure 3. The physical reasons are as follows. First, the $^{56}$Ni yields of the outflows must decrease as the weak outflows ($p \leq 0.3$) increase the density and temperature of the disk. Second, the area below the regions of Figure 3 corresponding the $^{56}$Ni yields of CCSNe. In general, previous CCSNe studies \citep[e.g.,][]{Heger2003,Fryer2004} often find that the $^{56}$Ni yields are several to tens of percents of solar mass. Nevertheless, the contributions of the strong outflows of NDAFs should still be highlighted.

It is obvious that the contributions of NDAFs to the $^{56}$Ni yields mainly reflect the progenitor stars in the range of $25-50~M_\odot$, which is comparable to the CCSNe of similar masses. In the previous work \citep{Song2019}, we proposed that the NDAF outflows are sufficient to power all observed SNe associated with GRBs; however, we did not consider the BH mass effect. This effect is considered in this work. We find that the total yields are still enough to explain all SN-GRB events, including the luminous ones. Hence, the nucleosynthesis of NDAFs with outflows might be responsible for resolving the crisis of the $^{56}$Ni yields in the luminous SNe. The evolution of the BH mass and spin in the hyperaccretion process might be further considered, especially in the faint explosions including fallback SNe \citep{Wei2021b}. Moreover, we find that the total $^{56}$Ni yields are insensitive to the progenitor masses. In many previous works, the total $^{56}$Ni yields are often obtained by fitting the CCSNe lightcurves. That is, the progenitor mass, and even the metallicity, are constrained by assuming that the $^{56}$Ni yields are solely originated from CCSNe. This assumption seems to be invalid since the nucleosynthesis of NDAF with outflows can produce a considerable amount of $^{56}$Ni. Hence, it might be inappropriate to infer the progenitor properties from the $^{56}$Ni yields. Additional observational evidences (along with the lightcurves) should be adopted to possibly distinguish the properties of progenitor stars.

\section{Applications on iron products}

The CCSN lightcurves are mainly driven by the decay of radioactive $^{56}$Ni and its daughter $^{56}$Co to $^{56}$Fe within the half-lives about 6.077 days and 77.236 days, respectively.
In most cases of NDAFs with outflows or CCSNe, the yields of $^{56}$Ni and its isotopes are generally larger than these of $^{56}$Fe and its isotopes. The products from the decay of
$^{56}$Ni in the collapsars should be revisited to include the nucleosynthesis of NDAFs with outflows.

There are many sophisticated chemical evolution models, considering the CCSNe relevant chemical enrichments \citep[see, e.g.,][]{DeDonder2003,Kobayashi2020}. However, all of them have neglected the possibility of the chemical enrichment through the NDAF outflows in the faint explosions.
Therefore, in order to highlight the importance of the nucleosynthesis in the NDAF outflow as an extra chemical enrichment source, we consider a simple closed-box chemical evolution model, which shall be validated by the current state-of-the-art chemical evolution model of \citet[][private communication]{Kobayashi2020}.

Considering a closed-box model for the chemical evolution of the cold gas of star-forming galaxies \citep[e.g.,][]{Tinsley1980,Matteucci2012book}, we have the following equations describing evolutions of the star formation, the gas, and the abundance of the element $i$ of interest:
\beq
\dot M_{\rm s} ~&=&~ M_{\rm g} ~ / ~ t_\star,\\
\frac{d M_{\rm g}}{dt} ~&=&~ - \dot M_{\rm s} + R,\\
\frac{d(M_{\rm g} X_{i})}{dt} ~&=&~ - \dot M_{\rm s} X_{i} + E_{i},
\eeq
where $\dot M_{\rm s}$ is the star formation rate, $M_{\rm g}$ is the gas mass, $t_\star$ is a typical star formation timescale, $R$ is the mass ejection rate at the end of stellar evolution, $X_i$ is the mass fraction of the element $i$, and $E_{i}$ is the ejection rate of the element $i$.

\begin{figure}
\centering
\includegraphics[width=1.0\linewidth]{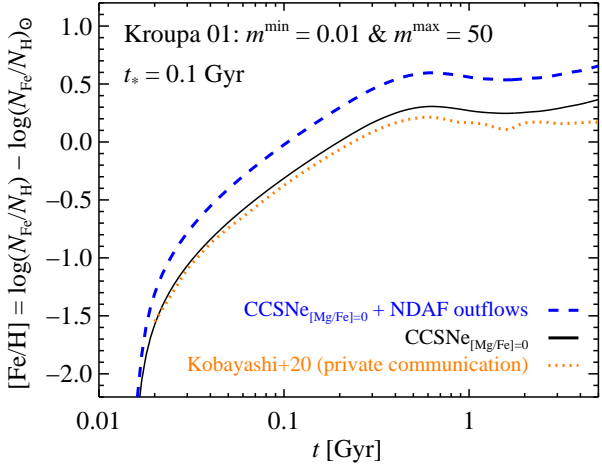}
\caption{Metallicity evolutions of $^{56}$Fe in cases of CCSNe combining with (blue dashed line) or without (black solid line) NDAF outflows by using the medians in Figure 3. The implication of the state-of-the-art chemical evolution model of \citet[][orange dotted line; private communication]{Kobayashi2020} for a close-box case without NDAF outflows is illustrated for comparison as well.}
\end{figure}

The gas is not only consumed by the current star formation but also restored by previously formed stars at the end their lives according to
\beq
R(t) = \int_{m_{\star}(t)}^{m^{\rm max}} (m - m_{\rm rem}) ~ \phi(m) ~ \dot M[t - \tau(m)] dm,
\eeq
where $m$ is the stellar mass in units of solar mass, $m_{\rm rem}(m)$ is the remnant mass, $\tau(m)$ is the lifetime of stars of mass $m$, and $m_\star$ is the mass of the star for which
$\tau(m_\star) = t$. To better compare with the model in \citet{Kobayashi2020}, we adopt the same \citet{Kroupa2001} initial mass function (IMF) for the number of stars, $dN$, within $d m$ as $\phi(m) \equiv dN/dm \propto m^{-\alpha_i}$ with $\alpha_1 = 0.3$ for $m^{\rm min} = 0.01 \leq m \leq 0.08$, $\alpha_2 = 1.3$ for $0.08 \leq m \leq 0.5$, and $\alpha_3 = 2.3$ for $0.5 \leq m \leq 50 = m^{\rm max}$, and it is normalized as $\int^{m^{\rm max}}_{m^{\rm min}} m \phi(m) dm = 1$. The remnant mass is given by \citep[e.g.,][]{Weidemann1983,Iben1984,Thorsett1999,Pagel2009book}
\beq
m_{\rm rem} \approx
\begin{cases}
0.106 m + 0.446, & 0.5 < m < 9\\
1.4, & 9 < m < 25 \\
0.24 m -4, & m > 25
\end{cases}
\eeq
while the lifetime of a star of mass $m$ with solar metallicity is from the work of the Geneva group \citep{Schaller1992} and approximated by
\beq
\tau(m) \approx 11.3 m^{-3} + 0.06 m^{-0.75} + 0.0012~~{\rm Gyr}.
\eeq

The total amount of the element $i$ ejected from stars is
\beq
E_{i}(t) \simeq \int_{m_{\star}(t)}^{m^{\rm max}} & \{ (m - m_{\rm rem}) X_i[t - \tau(m)] + q_{i}(m) \}
\nonumber\\
& \times~\phi(m)~\dot M[t - \tau(m)]~dm,
\eeq
where $q_{i}(m)$ is the fresh stellar yield of the element $i$.

We specifically consider the production of the element $^{56}$Fe totally decayed by $^{56}$Ni, i.e., $i~=~^{56}$Fe. For an amount of initial metal-free gas with mass $M_{\rm g,0}$ and star formation timescale $t_\star = 0.1~\rm Gyr$, assuming the median $^{56}$Fe yields as shown in Figure 3 {to stress the NDAF outflow contribution, we show the mass evolutions of $^{56}$Fe with and without the contribution of the NDAFs in Figure 4.

We also illustrate the $^{56}$Fe evolution implied by the \citet[][private communication]{Kobayashi2020} model for the closed-box case in Figure 4. Interestingly, our simple treatment agrees quite well with \citet{Kobayashi2020}, while the small difference may be attributed to the more accurate recipe adopted for the stellar lifetimes, i.e., the metallicity-dependent, and/or that developed for the delicate chemical enrichments by \citet{Kobayashi2020}.

Nevertheless, comparing to the CCSNe in the condition of [Mg/Fe]=0, the ratio of $^{56}$Fe mass to the initial total gas mass still could increase by a factor of $\simeq 1.95$ if the NDAFs are considered. Our results depend weakly upon the choice of the popular IMFs, such as those preferred by \cite{Cai2020}. This is because these popular IMFs often give similar ratios of the $8\ M_{\odot}\lesssim M_{\mathrm{pro}}\lesssim 30\ M_{\odot}$ stars to the heavier ones. Note that the increase factor is indeed in close relationship to the adopted maximum stellar mass, $m^{\rm max}$, in normalizing the IMF. For example, the increase factor decreases from $\simeq 1.95$ to $\simeq 1.55$ if $m^{\rm max}$ increases from 50 $M_\sun$ to 100 $M_\sun$, owing to the deficiency of the $^{56}$Fe enrichment through the NDAF outflow in stars massive than 50 $M_\sun$ (Figure 3).

Just like CCSNe, NDAFs can also produce $\alpha$ elements. We then estimate the cosmic buildup history of iron and $\alpha$ elements in the presence of NDAFs. That is, $\alpha$ elements are produced by CCSNe and NDAFs; SNe Ia, CCSNe and NDAFs contribute to the iron element. Our estimation procedures are as follows \citep[e.g.,][]{Blanc2008,Graur2015,Maoz2017}. First, we estimate the SN Ia rate as a function of redshift by adopting the cosmic star formation history (SFH) of \cite{Madau2017} and the delay-time distribution of SNe Ia of \cite{Maoz2017}. Second, we adopt the mean iron yield of a SN Ia as $y_{\mathrm{Ia}}=0.7\ M_{\odot}$ \citep[e.g.,][]{Howell2009,Graur2015,Maoz2017}. Third, for the mean iron yield of CCSNe $y_{\mathrm{CCSN}}$, we do not use the solid lines in Figure~3 since it corresponds to a optimistic situation; instead, we follow \cite{Maoz2017} and assume $y_{\mathrm{CCSN}}=0.074\ M_{\odot}$. Forth, for the \cite{Kroupa2001} IMF, the CCSN rate (the NDAF rate is assumed to be the same as the CCSN rate) is simply $0.01$ times the cosmic SFH. Fifth, the mean iron yield of NDAFs is estimated by considering the results in Figure 3 and the \cite{Kroupa2001} IMF. That is, the only difference between our calculation and that of \cite{Maoz2017} is that we take the nucleosynthesis of NDAFs with outflows into consideration. We then evaluate the volumetric iron-mass density $\rho_{\mathrm{tot}}(z)$ as a function of redshift, i.e., $\rho_{\mathrm{tot}}(z) = \rho_{\mathrm{Ia}}(z)+\rho_{\mathrm{CCSN}}(z)+\rho_{\mathrm{NDAF}}(z)$, where $\rho_{\mathrm{Ia}}(z)$, $\rho_{\mathrm{CCSN}}(z)$, and $\rho_{\mathrm{NDAF}}(z)$ are the densities due to SNe Ia, CCSNe, and NDAFs, respectively. Following Equations (5-8) in \cite{Maoz2017}, the $\alpha$-to-iron abundance ratio is
\begin{equation}
[\alpha/\mathrm{Fe}] (z) =\log \frac{f_{\rm CCSN}(z) + f_{\rm NDAF}(z)}{f_{\rm CCSN}(z=0.43) + f_{\rm NDAF}(z=0.43)},
\end{equation}
where $f_{\rm CCSN}=\rho_{\mathrm{CCSN}}/\rho_{\mathrm{tot}}$ and $f_{\rm NDAF}=\rho_{\mathrm{NDAF}}/\rho_{\mathrm{tot}}$. Note that the lookback time of redshift $z=0.43$ corresponds to the age of the Sun. Our result is shown in Figure~5 which displays the cosmic evolution of $[\alpha/\mathrm{Fe}] (z)$ with or without the contribution of NDAFs. In the presence of NDAFs, $[\alpha/\mathrm{Fe}] (z)$ is less sensitive to redshift than that without NDAFs. This is simply because the relative contribution to the iron production of SN Ia (i.e.,
$\rho_{\mathrm{Ia}}/\rho_{\mathrm{tot}}$) decreases considerably at redshift $z<0.5$ by considering the nucleosynthesis of NDAFs. Our results might contribute to the explanation of the lack of the cosmic evolution of the flux ratio of Fe II to Mg II in AGNs from low to high redshifts \citep[e.g.,][]{Barth2003,Jiang2007,Shin2019}.

\begin{figure}
\centering
\includegraphics[width=1.0\linewidth]{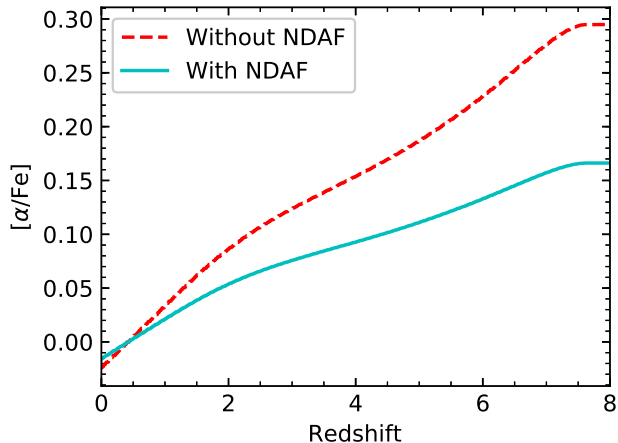}
\caption{The cosmic evolution of $[\alpha/\mathrm{Fe}] (z)$ with (solid line) or without (dashed line) the contribution of NDAFs.}
\end{figure}

\section{Conclusions and discussion}

We studied the NDAFs with outflows in the collapsar scenario and presented their contributions to the nucleosynthesis in scenes of the CCSNe and galaxies. The main conclusions are as follows.

(i) NDAFs are not only the GRB central engines but also the MeV neutrinos and GWs sources and the nucleosynthesis factories in the center of the collapsars or compact object mergers.

(ii) The nucleosynthesis of NDAFs with outflows is an important supplement to CCSNe. The exact composition of these outflows remains a subject of study in detailed models \citep[see, e.g.,][]{Zenati2020,Miller2020}. By considering the contributions of the NDAFs on the $^{56}$Ni yields, the lightcurves of CCSNe associated with GRBs could be well explained. In addition, our results indicate that it may be inappropriate to infer the progenitor properties from the $^{56}$Ni products.

(iii) As well as the hypernova models \citep[e.g.,][]{Kobayashi2020}, the yields of $^{56}$Fe decayed by $^{56}$Ni from the NDAF outflows in the center of the faint or failed CCSNe should be considered in the chemical evolution of galaxies and AGNs, which might help to understand the metal abundance in the solar neighborhood and the high-metallicity quasars in high redshift.

The energetic CCSN profiles in the simulations of \citet{Umeda2008} and the data of the final BH mass in \citet{Heger2002} are chosen in order to describe the possibly significant contribution of the NDAF outflows on nucleosynthesis in this work. Actually, there are amount of increasingly sophisticated simulations on CCSNe and their nucleosynthesis processes \citep[e.g.,][and references therein]{Nakamura2001,Maeda2003,Fujimoto2007}, which might be referred to bring more precise results on the heavy-element productions in the uniform description of the stellar evolution. Recently, we built the 1D numerical simulations on CCSNe by Athena++ to investigate the first (or lower) mass gap in the compact object mass distribution by considering the progenitors with different mass, metallicity, and initial explosion energy \citep{Liu2021}. The more self-consistent solutions on nucleosynthesis of CCSNe including NDAF outflows will be done in our future works.

\acknowledgments
We thank Prof. Chiaki Kobayashi for private communication and the anonymous referee for helpful suggestions and comments. This work was supported by the National Natural Science Foundation of China under grants 11822304, 11890693, and 11973002, and the science research grants from the China Manned Space Project with No. CMS-CSST-2021-B11.

\end{document}